\pdfoutput=1
\documentclass[a4paper,12pt]{article}
\usepackage{cite}
\usepackage{amsmath,amssymb,amsfonts}
\usepackage[multiple,stable]{footmisc}
\usepackage[amsmath,amsthm,thmmarks]{ntheorem}
\allowdisplaybreaks[4]
\usepackage[noconfig]{refstyle}
\usepackage[dvipsnames]{xcolor}
\usepackage[hyperfootnotes=false, linktocpage=true, colorlinks, citecolor=blue, linkcolor=blue, urlcolor=Maroon]{hyperref}
\usepackage{array,tabularx,booktabs,geometry,subfigure,graphicx,longtable}
\usepackage[bf]{caption}
\usepackage{tikz}
\usepackage{multirow}
\usetikzlibrary{shapes,arrows}
\tikzstyle{block} = [rectangle, draw, text width=7em, text centered, rounded corners, minimum height=3em]

\let\eqref=\relax
\newref{f}{name={footnote~},Name={Footnote~},names={footnotes~},Names={Footnotes~}}
\newref{app}{name={appendix~},Name={Appendix~},names={appendixes~},Names={Appendixes~}}
\newref{tab}{name={table~},Name={Table~},names={tables~},Names={Tables~}}
\newref{ch}{name={chapter~},Name={Chapter~},names={chapters~},Names={Chapters~}}
\newref{sec}{name={section~},Name={Section~},names={sections~},Names={Sections~}}
\newref{fig}{name={figure~},Name={Figure~},names={figures~},Names={Figures~}}
\numberwithin{equation}{section}

\newcommand{\be}{\begin{equation}}
\newcommand{\ee}{\end{equation}}
\newcommand{\bea}{\begin{equation}\begin{aligned}}	
\newcommand{\eea}{\end{aligned}\end{equation}}		

\newcommand{\iddots}{\mathinner{\mkern2mu\raise1pt\hbox{.}\mkern2mu \raise4pt\hbox{.}\mkern2mu\raise7pt\hbox{.}\mkern1mu}}

\providecommand{\id}{\leavevmode\hbox{\small$\mathrm{1}$\kern-3.8pt\normalsize$\mathrm{1}$}}
\def\fnote#1#2{\begingroup\def\thefootnote{#1}\footnote{#2}
     \addtocounter{footnote}{-1}\endgroup}

\begin{document}

\vspace{1cm}

\title{
       \vskip 40pt
       {\Large \bf Structure In The Heterotic \\ Matter Field K\"ahler Potential}}

\vspace{2cm}
\author{James Gray }
\date{}
\maketitle
\begin{center} {\small {\it Physics Department, Robeson Hall, Virginia Tech,\\ Blacksburg, VA 24061, U.S.A.}}\\
\fnote{}{jamesgray@vt.edu}
\end{center}

\begin{abstract}
It has long been known that superpotential couplings in heterotic theories often vanish despite the absence of any known symmetry that would forbid them. We show that such structure is also exhibited by the matter field K\"ahler potential of these compactifications. We give an analytical analysis which provides some simple conditions under which certain terms, which would be expected to be present in the K\"ahler potential, vanish. Explicit examples of this phenomenon are provided, some of which are verifiable by other methods and some of which can only currently be accessed by these techniques. By studying the interplay of the structure we find with Higgsing transitions, we are able to show that this vanishing of K\"ahler potential terms can extend to all orders in a perturbative expansion in the matter fields. 
\end{abstract}

\thispagestyle{empty}
\setcounter{page}{0}
\newpage

\tableofcontents


\section{Introduction}

It has long been known that, in the context of smooth Calabi-Yau compactifications of the heterotic string, superpotential couplings often vanish for no obvious four-dimensional reason \cite{Braun:2006me,Bouchard:2006dn,Anderson:2010tc,Buchbinder:2014sya,Blesneag:2015pvz,Blesneag:2016yag,Gray:2019tzn,Anderson:2021unr,Anderson:2022kgk,Gray:2024xun} (see \cite{Kobayashi:2011cw,CaboBizet:2013gns,Parameswaran:2014xla} for examples of some related work in an orbifold context). That is, there are frequently couplings which are consistent with all of the known symmetries of the four dimensional effective theory, which simply vanish when computed in an explicit construction. Initial work in this area concerned vanishing of perturbative tri-linear Yukawa couplings \cite{Strominger:1985ks,Greene:1986bm,Greene:1986jb,Candelas:1987se,Distler:1987gg,Distler:1987ee,Greene:1987xh,Candelas:1987rx,Candelas:1990pi,Distler:1995bc,Berglund:1995yu,Braun:2006me,Bouchard:2006dn,Anderson:2010vdj,Anderson:2010tc,Buchbinder:2014sya,Blesneag:2015pvz,Blesneag:2016yag,McOrist:2016cfl,Ashmore:2018ybe,Gray:2019tzn,Anderson:2021unr,Anderson:2022kgk,Ibarra:2024hfm}. More recent progress has observed that such structure extends to all higher order perturbative couplings of the matter fields \cite{Gray:2024xun}. This latter work leads to infinite numbers of vanishing perturbative couplings in some cases. Vanishing results of this type have also been found for non-perturbative contributions to the superpotential as well \cite{Buchbinder:2019eal,Buchbinder:2019hyb,Buchbinder:2018hns,Buchbinder:2017azb,Buchbinder:2016rmw,Bertolini:2014dna,Beasley:2003fx,Basu:2003bq,Buchbinder:2002pr,Buchbinder:2002ic,Lima:2001jc,Witten:1999eg,Harvey:1999as,Witten:1996bn,Dine:1987bq,Distler:1987ee,Dine:1986zy,Distler:1986wm}, although it has been conjectured that the vanishing of non-perturbative effects will be limited in a given case except in certain very specific circumstances \cite{Palti:2020qlc}.

The vanishing of perturbative terms in the superpotential, as described in the previous paragraph, is interesting even if those terms are regrown non-perturbatively at some level. If all that was known  were the quantum numbers of the four dimensional massless fields, and the known symmetries of the effective theory, then these results would come as a surprise. From a field theory point of view, it should not be possible, for completely general choices of numerical coefficients in a theory, to find a region of field space where the number of suppressed couplings, not enforced by symmetry, is larger than the number of moduli. However, this is precisely what is found in some of these string constructions.

A natural question to ask is whether this phenomenon of vanishing terms in the low energy description extends to the matter field K\"ahler potential of the heterotic theory. In this paper we show that it does. We provide simple criteria, that can be evaluated analytically, for when perturbative terms, quadratic in the matter fields, vanish. These terms are often consistent with all known symmetries and would therefore be expected to appear in the four dimensional K\"ahler potential. We give a series of explicit examples illustrating this structure, including some that can be verified by known symmetries and some that can not.

One might ask if such vanishing components of the matter field metric can be meaningful. Indeed, by a field redefinition it should always be possible to diagonalize that metric and it must be of maximal rank. As such, in some sense one might think that no good notion of vanishing components could survive. However, the matter field metric is a complicated function of the moduli of the compactification. Although a field transformation certainly exists to diagonalize the metric, it will not in general be a holomorphic function of the $N=1$ degrees of freedom. In this sense, the structure we observe is indeed physically meaningful.

This work can be seen as a complement to recent numerical explorations of the heterotic matter field K\"ahler potential \cite{Constantin:2025vyt,Butbaia:2024xgj,Berglund:2024uqv,Constantin:2024yxh,Butbaia:2024tje} (see \cite{Gray:2003vw,Blesneag:2018ygh} for other analytic work on computing gauge sector K\"ahler potentials). Such numerical methods are absolutely necessary currently, if we wish to compute the non-zero numbers that appear in this quantity. However, the current work shows that certain structure, in particular vanishing of terms, in the K\"ahler potential can be obtained analytically.

Surprisingly, in the last section of the paper we will be able to make some statements about vanishings of higher than quadratic order perturbative terms in the matter field K\"ahler potential. This will be made possible, despite the fact that no tractable closed form expressions for the coefficients of such terms are available, by considering how the constraints on K\"ahler potentials we have described here interact with Higgsing transitions. In this manner, we will be able to show that infinite sets of perturbative contributions to the matter field K\"ahler potential vanish in some examples: an analogous result to those obtained for the superpotential in \cite{Gray:2024xun}.

An underlying question which this work doesn't address is whether there is a, thus far undiscovered, symmetry that underlies these vanishings, perhaps along the lines of \cite{Palti:2020qlc}, or whether this structure is simply a geometrical feature of these compactifications. This is a subject that we plan to return to in future work.

The rest of this paper is structured as follows. In Section \ref{math}, some of the underlying mathematics that will be needed for our discussion is reviewed. In Section \ref{vansec} the basic result detailing vanishings of terms in heterotic matter field K\"ahler potentials is outlined. A series of explicit examples are presented in Section \ref{sec:eg}, each emphasizing a different point. The first example has vanishing structure which is verifiable by symmetry considerations. The second example has vanishing structure which is not explained by known symmetries. The final example is used to argue that the vanishing phenomenon being discussed extends to higher order terms in the matter field K\"ahler potential. In Section \ref{sec:conc} we conclude and mention some examples of possible future directions of research.

\section{Hermitian metrics and maps on cohomology} \label{math}

In this section we will review a few mathematical preliminaries that will be required in the rest of our discussion. See, for example, \cite{huybrechts} or Appendix C of \cite{Blesneag:2015pvz} for a more detailed review of some of these notions.

Consider a short exact sequence of bundles,
\begin{eqnarray} \label{mrmon}
0 \to A \stackrel{i}{\longrightarrow} V \stackrel{m}{\longrightarrow} B \to 0 \;.
\end{eqnarray}
Holomorphic maps such as $i$ and $m$ induce well defined maps in cohomology, which we will indicate by the same symbol in a slight abuse of notation. It should be noted, however, that as maps on forms, holomorphic maps such as $i$ do not preserve harmonicity, as we will now review. 

A form valued in the bundle $E$ is said to be harmonic if,
\begin{eqnarray}
 \overline{\partial}_{E} \alpha=0 \;\;\; \text{and} \;\;\; \overline{\partial}_{E}^{\dagger} \alpha=0\;.
\end{eqnarray}
Here, if we write $\alpha=\alpha^i \otimes s_i$ where $(s_1,s_2,\ldots)$ is a local holomorphic trivialization, the partial derivative operator is defined by,
\begin{eqnarray}
    \overline{\partial}_E \alpha = \overline{\partial}{\alpha^i} \otimes s_i
\end{eqnarray} 
The adjoint of the partial derivative operator is defined by $\overline{\partial}^{\dagger}_{E} \alpha= - \overline{\star}_{E^{\vee}} \circ \overline{\partial}_{E^{\vee}} \circ \overline{\star}_{E}$. In this expression we have assumed the existence of a hermitian structure $h_E$ which can be thought of as a map taking $E\to E^{\vee}$. This is then used to define the generalization of the Hodge duality operator which is being used in the above.
\begin{eqnarray}
    \overline{\star}_E(\alpha^i \otimes s_i) = \star(\overline{\alpha}^i) \otimes h_{E \,ij} s^j
\end{eqnarray}

Although an holomorphic map such as $i$ in (\ref{mrmon}), for example, obviously maps $\overline{\partial}_A$ closed and exact forms to objects of the same type, it does not preserve $\overline{\partial}^{\dagger}$ closedness. Taking $\omega^h$ as a harmonic $A$ valued form,
\begin{eqnarray}
\overline{\partial}^{\dagger}_V i \omega^h = \overline{\star}_{V^{\vee}} \circ \overline{\partial}_{V^{\vee}}\circ  \overline{\star}_{V} i \omega^h \neq 0 \;.
\end{eqnarray}
Here the last inequality follows due to $\overline{\star}_{V}$ conjugating the holomorphic map $i$ which is then acted upon by $\overline{\partial}_{V^{\vee}}$. A generalization of the Hodge decomposition states that every $V$ valued $(p,q)$ form can be written as a sum 
\begin{eqnarray} \label{hodge}
\alpha = \alpha_h +\overline{\partial}_V \beta + \overline{\partial}^{\dagger}_V \gamma
\end{eqnarray} 
in a unique fashion, where $\alpha_h$ is harmonic. The fact that $i \omega_h$ is $\overline{\partial}_V$ closed but not $\overline{\partial}^{\dagger}_V$ closed then implies from (\ref{hodge}) that $i\omega_h = \alpha_h + \overline{\partial}_V \beta$ for some $\alpha_h$ and $\beta$. It is, of course, still possible to induce a map on harmonic forms from $i$. One simply considers the well defined map on cohomology it induces and then takes the unique harmonic representative $\alpha_h$ in any class mapped to.

Given hermitian structures on the bundles in the extension (\ref{mrmon}), one can define maps $i^*: \Omega^{p,q}(V) \to \Omega^{p,q}(A)$ and $m^*: \Omega^{p,q}(B) \to \Omega^{p,q}(V)$ on bundle valued forms. We have,
\begin{eqnarray}
    i^* = \overline{\star}_{A^{\vee}} \circ i^{\vee} \circ \overline{\star}_{V} \;\;\; \text{and} \;\;\;m^* = \overline{\star}_{V^{\vee}} \circ m^{\vee} \circ \overline{\star}_{B} \;.
\end{eqnarray}
These maps induce well defined maps on bundle valued cohomology, which we will again denote by the same symbol, simply by composition. The Hodge stars and holomorphic maps from which they are built provide us with such morphisms and thus so do their sequential application. Note also that,
\begin{eqnarray} \label{composition}
    i^* \circ m^*= \overline{\star}_{A^{\vee}} \circ i^{\vee} \circ \overline{\star}_V \circ \overline{\star}_{V^{\vee}} \circ m^{\vee} \circ  \overline{\star}_B =  (-1)^{p+q}\overline{\star}_{A^{\vee}} \circ i^{\vee}\circ m^{\vee} \circ  \overline{\star}_B =0 \;,
\end{eqnarray}
given that $\overline{\star}_{E}\circ \overline{\star}_{E^{\vee}}=(-1)^{p+q}$ and $i^{\vee} \circ m^{\vee}=0$.

As with the holomorphic maps $i$ and $m$, $i^*$ and $m^*$, as maps on differential bundle valued forms, do not preserve harmonicity. In particular, taking $\nu^h$ as a $B$ valued harmonic form, given that
\begin{eqnarray}
    m^* \nu^h =  \overline{\star}_{V^{\vee}} \circ m^{\vee} \circ \overline{\star}_{B} \;\nu^h \;,
\end{eqnarray}
we see that, because $\overline{\star}_{B} \;\nu^h$ is harmonic, $m^{\vee} \circ \overline{\star}_{B} \;\nu^h$ is harmonic up to an exact piece, following the same logic as we did for the map $i$ above. Thus, because $\overline{\star}_{V^{\vee}} \circ \overline{\partial}_{V^{\vee}} = (-1)^{p+q+1}\;\overline{\partial}^{\dagger}_{V} \circ \overline{\star}_{V^{\vee}}$, we find that $m^* \nu^h$ is harmonic up to a $\overline{\partial}^{\dagger}_V$-exact piece . This same relation also implies that the harmonic part of the image of the map $m^*$ in forms is the same as the harmonic representative given by the induced map in cohomology.

Maps such as $m^*$ have a very important property with respect to the inner product defined on bundle valued cohomology.
\begin{eqnarray} \label{inner}
\left<\alpha, \beta \right>_V = \int_X \overline{\star}_V \alpha \wedge \beta
\end{eqnarray}
In this expression, the dual bundle valued form $\overline{\star}_V \alpha$ acts on the bundle valued form $\beta$ to produce an ordinary top form which can be integrated over the Calabi-Yau $X$ to yield a number. A straight forward calculation reveals that $m^*$, as a map on forms, obeys the following adjoint property where $\alpha$ is a $(p,q)$ form.
\begin{eqnarray} \label{reln1}
\left<m^*\alpha, \beta \right>_V = (-1)^{p+q}\left< \alpha , m\beta\right>_B
\end{eqnarray}
In the physical application of this paper, we are interested in inner products between harmonic forms. One might therefore worry that, because $m^* \alpha$ as defined above is not harmonic, the above relation does not apply if we are thinking of $m^*$ as a map on cohomology and from there on harmonic objects. This is not an issue however. If $\alpha$ is harmonic, the form $m^* \alpha$ differs from the relevant harmonic form by a co-exact piece. The derivative $\overline{\partial}^{\dagger}_V$ is adjoint to $\overline{\partial}_V$ and thus the addition of this co-exact piece does not change the inner product given that $\beta$ is closed. Likewise, $m\beta$ differs from the harmonic representative in cohomology by an exact piece. This does not affect the second inner product in (\ref{reln1}) if we take $\alpha$ to be $\overline{\partial}^{\dagger}_V$ closed.

\section{Vanishing of terms in the matter field K\"ahler potential} \label{vansec}
Using the discussion of the previous section, let us state the basic vanishing structure we will consider in this paper. Consider a heterotic Calabi-Yau compactification with gauge bundle $V$. If we have matter fields $f^i$ associated to bundle valued one forms $\omega_i$ then the matter field kinetic terms in the Lagrangian take the form 
\begin{eqnarray}
  G_{i \overline{j}} \partial_{\mu} f^i \partial^{\nu} \overline{f}^{\overline{j}} \;\; \text{where} \;\; G_{i \overline{j}} \propto \left< \omega_j, \omega_i\right>_V
\end{eqnarray} 
That is, the components of the matter field K\"ahler metric in a four dimensional effective theory of a heterotic compactification are determined precisely by the inner product (\ref{inner}).

Say that our bundle $V$ is defined by an extension short exact sequence as in (\ref{mrmon}) where $A$ and $B$ are bundles\footnote{Very closely analogous analyses can be made for other bundle constructions, such as monads.}. This defining sequence leads to a long exact sequence in cohomology, which includes a piece of the following form.
\begin{eqnarray} \label{lespiece}
H^0(B) \stackrel{\delta}{\to} H^1(A) \stackrel{i}{\to} H^1(V) \stackrel{m}{\to} H^1(B) \stackrel{\delta}{\to} H^2(A)
\end{eqnarray}
In this sequence, $\delta$ is the co-boundary map and, in a slight abuse of notation, we have denoted induced maps in cohomology by the same symbol as the associated bundle morphism. Consider two matter fields corresponding to $V$ valued harmonic forms $m^*\beta$ and $i \alpha$ respectively, for $\beta\in H^1(B)$ and $\alpha \in H^1(A)$. We then have the following.
\begin{eqnarray} \label{mrvan}
    \left<m^*\beta, i \alpha \right>_V =  -  \left<\alpha, m \circ i \beta \right>_B =0
\end{eqnarray}
Here we have used (\ref{reln1}) in the first equality and the composition rule for the maps in the sequence (\ref{lespiece}) in the second. The inner product in (\ref{mrvan}) is precisely that which computes the matter field K\"ahler metric component linking the fields associated to $m^*\beta$ and $i\alpha$. Thus, in such an instance, the kinetic cross-term between these two species vanishes. Equivalently, the associated moduli dependent terms $G_{i \overline{a}} f^i \overline{\tilde{f}}^{\overline{a}}$ in the matter field K\"ahler potential, for fields $f$ and $\tilde{f}$ associated to elements of the form $i\alpha$ and $m^*\beta$ respectively, vanishes. Note that the conjugate metric element also vanishes as it should,
\begin{eqnarray}
 \left< i \alpha ,m^*\beta\right>_V =  - \left<\alpha, i^* \circ m^* \alpha \right>_B =0
\end{eqnarray}
due to the compositional relationship (\ref{composition}).

\vspace{0.2cm}

One issue that it is useful, although not necessarily essential, to clarify in interpreting these results is whether or not harmonic forms of the types $i \alpha$ and $m^* \beta$ form a basis of $H^1(V)$. In fact, it is relatively straightforward to see that they do by following a somewhat indirect argument. Elements of the form $i \alpha$ span a subspace of $H^1(V)$ of dimension $\dim (\text{coker} [\delta:H^0(B) \to H^1(A)])$. This is easy to see because the kernel of $i$ is the image of $\delta$ by the exactness of the sequence (\ref{lespiece}). For the $m^* \beta$ elements consider the following sequence that follows from the dual sequence to (\ref{mrmon}).
\begin{eqnarray} \label{lespiece2}
H^2(V^{\vee}) \stackrel{m^{\vee}}{\leftarrow} H^2(B^{\vee}) \stackrel{\delta^{\vee}}{\leftarrow} H^1(A^{\vee})
\end{eqnarray}
From the exactness of this sequence we see that elements of the type $m^{\vee} b$ span a subspace of $H^2(V^{\vee})$ of dimension $\text{dim} (\text{coker} [\delta^{\vee}:H^1(A^{\vee}) \to H^2(B^{\vee})])$. However any $b\in H^2(B^{\vee})$ is $\overline{\star}_B \gamma$ for some $\gamma \in H^1(B)$ since the Hodge star provides an isomorphism of cohomology groups\footnote{In fact, $\overline{\star}_B$ maps $H^1(B)$ to $H^{3,2}(B^{\vee})$. However, since for a Calabi-Yau threefold $\wedge^3 TX^{\vee} ={\cal O}$, we have that $H^{3,2}(B^{\vee}) = H^2(\wedge^3 TX^{\vee} \otimes B^{\vee}) = H^2({\cal O}\otimes B^{\vee}) = H^2(B^{\vee})$. Similar comments apply for the other cohomologies of dual bundles being considered in this paragraph.}. Likewise, $\overline{\star}_{V^{\vee}}$ provides an isomorphism from $H^2(V^{\vee})$ to $H^1(V)$. Thus elements of the form $m^* \beta$ span a $\text{dim}(\text{coker}[\delta^{\vee}:H^1(A^{\vee}) \to H^2(B^{\vee})])=\text{dim}( \text{ker}[\delta:H^1(B) \to H^2(A)])$ dimensional subset of $H^1(V)$. Since $h^1(V) =\dim (\text{ker} \left[\delta:H^1(B) \to H^2(A)\right])+ \dim (\text{coker} \left[ \delta:H^0(B) \to H^1(A)\right])$ we naively have the right number of basis elements. The only obstruction to having a complete basis that could arise would be if some element $m^*\beta$ was actually the same as some element $i \alpha$ of $H^1(V)$. This can not occur however, precisely due to the vanishing result (\ref{mrvan}). All inner products of the form $\left<m^* \beta, i \alpha \right>$ vanish. However, if $m^* \beta =i \alpha$ for some $\alpha$ and $\beta$ then this would imply that $\left<\omega,\omega\right>=0$ for that $\omega=m^*\beta$, which can not happen due to the positivity properties of the inner product. Thus elements of $H^1(V)$ of the form $i\alpha$ and $m^* \beta$ form a basis of the cohomology.

\vspace{0.2cm}

Having established some vanishing statements about components of the matter field K\"ahler metric, we should address an obvious objection that could be raised. In some sense we know exactly how many vanishings there are in a metric such as this. Any metric can be diagonalized and any non-singular metric is of maximal rank. So one might think that the above results do not mean very much. There are two ways in which this is too quick. Firstly, the above discussion gives us some information concerning the basis in field space in which this diagonalization occurs. This information is incomplete and concrete computation of the map $f^*$ in a given example would require knowledge of the bundle and Calabi-Yau metrics. More importantly, the K\"ahler potential terms and corresponding matter field K\"ahler metric components that we are discussing here are functions of singlets such as bundle and complex structure moduli. Thus if one were to arrange for the vanishing structure we have seen by performing a basis change on the matter fields, this transformation would have to depend upon these moduli. Such a transformation would not be, however, holomorphic in these fields in general. As such it is incompatible with the $N=1$ supersymmetric structure imposed upon these singlet chiral superfields.

\vspace{0.2cm}

In this discussion we have considered $V$ valued cohomologies, however clearly similar analysis would hold for other associated bundles, such as wedge powers, corresponding to matter fields in different representations. In what follows, we will present a number of explicit examples of the above vanishing phenomenon. In some of these, the vanishings will be verifiable by using know symmetries of the string compactifications in questions. In others, no known symmetry results in the structure we are discussing. Finally, we shall present an example which demonstrates that there are many vanishings in higher order perturbative terms in the matter field K\"ahler potential as well, despite the fact that we have apriori no direct method to compute these quantities.

\section{Examples} \label{sec:eg}

\subsection{A vanishing example verifiable by symmetry}

We begin with an extremely simple example where the vanishing of K\"aher potential terms we are discussing can be verified by other means. Consider the following manifold,
\begin{eqnarray}
   X= \left[ \begin{array}{c|c} \mathbb{P}^1 & 2 \\ \mathbb{P}^1 & 2 \\\mathbb{P}^1 & 2 \\\mathbb{P}^1 & 2  \end{array}\right] \;,
\end{eqnarray}
and line bundle sum over it,
\begin{eqnarray} \label{sumeg}
   V={\cal O}(-2,2,0,0)\oplus {\cal O}(2,-2,0,0) \;.
\end{eqnarray}
Here we are using the standard notation for CICYs and line bundles over them, as reviewed, for example in \cite{Anderson:2008ex}. The bundle $V$ is holomorphic and slope-polystable in a locus in the K\"ahler cone of $X$ and thus corresponds to a good vacuum of the heterotic string. In this case we have a bundle of $S(U(1) \times U(1))$  structure giving rise to a low energy gauge group of $E7 \times U(1)$ where the $U(1)$ factor is Green-Schwarz anomalous. In terms of line bundle cohomology, it is easy to compute that $h^*({\cal O}(-2,2,0,0))=\left\{ 0,3,3,0 \right\}$ and thus $V$ in this case gives rise to 3 matter fields in the ${\bf 56}_1$ and 3 matter fields in the ${\bf 56}_{-1}$ representations of the gauge group.

To connect with our previous analysis, we can view (\ref{sumeg}) as a bundle defined by extension where the extension class is taken to be trivial.
\begin{eqnarray}
    \begin{array}{ccccccccc}
    0& \to &{\cal O}(-2,2,0,0)& \stackrel{i}{\to}& {\cal O}(-2,2,0,0)\oplus {\cal O}(2,-2,0,0) & \stackrel{m}{\to} & {\cal O}(2,-2,0,0)& \to& 0 \\ 
    h^0 & & 0 & &0&& 0\\
    h^1 &&3 &&6&&3 \\
    h^2 && 3 &&6 &&3 \\
    h^3 && 0&&0&&0\end{array}
\end{eqnarray}
For the purposes of a unified presentation with later examples, we have reproduced the dimensions of the cohomology groups of the bundles below their entry in the sequence.
In this case the map structure discussed in Section \ref{vansec} becomes particularly simple. The map $m$ is simply an isomorphism to $O(2,-2,0,0)$ from the relevant summand in $V$. In such a case $m^{\vee}$ is also an isomorphism onto the relevant summand in $V^{\vee}$ and thus $m^*$ is an isomorphism too. 

By the arguments of Section \ref{vansec} all matter field K\"ahler metric terms which would induce bilinear cross terms between the matter fields associated to $i \alpha$ and those associated to $m^*\beta$ vanish. In other words, denoting the matter fields associated to $i \alpha$ as $C$ and those associated to $m^* \beta$ as $\tilde{C}$ all terms in the matter field K\"ahler potential that are schematically of the form $\overline{C} \tilde{C}$ and its conjugate, with moduli dependent pre-factors, vanish.

In this particular case, this result can easily be confirmed. The matter fields $C$ correspond to the ${\bf 56}_1$ and the matter fields $\tilde{C}$ the ${\bf 56}_{-1}$ representations mentioned above. A term of the form $\overline{C} \tilde{C}$, with a singlet pre-factor, simply isn't invariant under the Green-Schwarz anomalous $U(1)$ symmetry. It is therefore forbidden in the perturbative K\"ahler potential. Note that such terms might be regrown non-perturbatively, in a suppressed fashion in regions of moduli space that are under control, due to the K\"ahler axions having their shift symmetry gauged under the same $U(1)$. 

\subsection{A vanishing example not due to any known symmetry} \label{eg2}

With the example in this section we wish to illustrate that not every vanishing of the type we have been discussing in this paper is due to a known symmetry, as happened to be the case in the proceeding example. Consider the following Calabi-Yau threefold,
\begin{eqnarray}
    X=\left[ \begin{array}{c|cc} \mathbb{P}^1 &0 &2\\\mathbb{P}^1 &2&0\\\mathbb{P}^3 &3&1\end{array} \right] \;,
\end{eqnarray}
together with the following $SU(3)$ bundle over it.
\begin{eqnarray} \label{v3other}
\begin{array}{ccccccccc}
    0& \to &{\cal F}& \stackrel{i}{\to}& V_3 & \stackrel{m}{\to} &{\cal K}& \to& 0 \\ 
    h^0 & & 0 & &0&& 0\\
    h^1 &&5 &&5&&0 \\
    h^2 && 3 &&7 &&4 \\
    h^3 && 0&&0&&0\end{array}
\end{eqnarray}
Here we have defined
\begin{eqnarray} \label{fdef}
    0\to {\cal O}(-1,-2,1) \to {\cal F} \to {\cal O}(-2,2,0) \to 0\;,
\end{eqnarray}
and 
\begin{eqnarray}
    {\cal K} = {\cal O}(3,0,-1)\;.
\end{eqnarray}
Our notation has been chosen to mirror that in Section 4.2 of \cite{Anderson:2010ty} the analysis of which has been used to construct this example and that used in the next section. It has been explicitly checked that $V_3$ is poly-stable and slope zero in regions of the K\"ahler cone relevant for the discussion of this paper. In (\ref{v3other})  we have once again included the dimensions of the relevant cohomology groups underneath the bundles in the sequence. We see that one obtains $4$ matter fields $\tilde{f}_1$ in the $\overline{\bf 16}$ representation from acting on elements of $H^2({\cal K})=H^1({\cal K}^{\vee})$ with $m^*$ and $3$, $\tilde{f}_2$, as the image under $i$ of an element of $H^2({\cal F})=H^1({\cal F}^{\vee})$. According to our analysis, the cross-terms of the form $\tilde{f_2} \overline{\tilde{f}}_1$ and their complex conjugates in the matter field K\"ahler potential vanish\footnote{Note that the vanishing of the inner product between these two forms following the analysis of Section \ref{vansec} trivially implies the vanishing of the inner product between the associated one forms in the dual first cohomology group.}. In this example, there is no obvious symmetry reason for this vanishing, because such K\"ahler potential terms are permitted by the $E_6$ gauge group which is the only relevant gauge symmetry of the problem.

One may ask if the vanishing of these matter field metric components is due to non-apparent symmetries stemming from this bundles obvious stability wall \cite{Sharpe:1998zu,Anderson:2009sw,Anderson:2009nt,Anderson:2010tc,Buchbinder:2014sya}. In particular, at the locus in the K\"ahler cone where ${\cal F}$ becomes slope zero, $V_3$ in (\ref{v3other}) must split into two pieces to maintain supersymmetry \cite{Sharpe:1998zu,Anderson:2009sw,Anderson:2009nt}. It becomes a bundle of structure group $S(U(2)\times U(1))$ of the following form.
\begin{eqnarray} \label{mrv3sum}
    V_3 \to {\cal F} \oplus {\cal K}
\end{eqnarray}

A bundle of this structure group leads to an $E_6\times U(1)$ gauge group in four dimensions. The $U(1)$ factor is Green-Schwarz anomalous. The relevant branching rule for the adjoint representation of $E_8$ in this situation is as follows.
\begin{eqnarray}\nonumber
       E_8 &\supset& E_6 \times SU(2) \times U(1) \\ 
    {\bf 248} &=&({\bf 1},{\bf 1})_0+({\bf 1},{\bf 2})_3+({\bf 1},{\bf 2})_{-3}+({\bf 1},{\bf 3})_0+({\bf 27},{\bf 1})_{-2}+({\bf 27},{\bf 2})_1 \\ \nonumber &&+(\overline{\bf 27},{\bf 1})_2+(\overline{\bf 27},{\bf 2})_{-1}+({\bf 78},{\bf 1})_{0}
\end{eqnarray}
This leads to a matter content for the theory at the stability wall as given in Table \ref{table:spec}.
\begin{table}[h]
\centering
\begin{tabular}{|c|c|c|c|}
\hline
Field & Representation & Cohomology & Dimension\\ [0.5ex] 
\hline
$C_1$&$({\bf 1},{\bf 2})_{-3}$& $H^1({\cal K}\otimes{\cal F}^{\vee})$ &0\\
$C_2$&$({\bf 1},{\bf 2})_3$&$H^1({\cal K}^{\vee}\otimes{\cal F}^{})$&20 \\
$f_1$&$({\bf 27},{\bf 1})_{-2}$&$H^1({\cal K})$& 0 \\
$f_2$ & $({\bf 27},{\bf 2})_{1}$ & $H^1({\cal F})$& 5\\
$\tilde{f}_1$ & $(\overline{\bf 27},{\bf 1})_2$ & $H^1({\cal K}^{\vee})$& 4\\
$\tilde{f}_2$ & $(\overline{\bf 27},{\bf 2})_{-1}$ & $H^1({\cal F}^{\vee})$& 3\\ 
$\phi$ & $({\bf 1},{\bf 3})_0$ & $H^1({\cal F}^{\vee} \otimes {\cal F})$& 23\\[1ex]
\hline
\end{tabular}
\caption{Matter content of the 4D effective theory arising from the split bundle (\ref{mrv3sum})}
\label{table:spec}
\end{table}

The four dimensional effective theory has D-terms associated to both factors in the gauge group. For the Green-Schwarz anomalous $U(1)$ factor we have the following.
\begin{eqnarray} \label{du1}
    D^{U(1)} = \frac{3}{16} \frac{\epsilon_S \epsilon_R^2}{\kappa_4^2} \frac{\mu({\cal F})}{\cal V}  + 3 {G}_{l \overline{m}}C_2^l \overline{C}_2^{\overline{m}}+ {\cal G}_{i \overline{j}} f_2^i \overline{f}_2^{\overline{j}} +2 \tilde{\cal G}_{I \overline{J}} \tilde{f}_1^I \overline{\tilde{f}}_1^{\overline{J}} 
\end{eqnarray}
Here we have only included terms in the D-term which will be relevant in this paper, omitting some contributions which will remain zero in all of the vacua we consider. In the expression (\ref{du1}), $\mu({\cal F})$ is the slope of ${\cal F}$, ${\cal V}$ is the Calabi-Yau volume (both functions of K\"ahler moduli) and the various forms of $G$'s are all portions of the matter field K\"ahler metric. The constants $\epsilon_S$, $\epsilon_R$ and $\kappa_4$ are all the usual constants appearing in the expansions controlling heterotic M-theory \cite{Lukas:1998hk}.

The D-term associated to the $E_6$ gauge group factor has the following schematic form.
\begin{eqnarray} \label{d2010}
    D^{SO(10)}=  {\cal G}_{i \overline{j}} f_2^i \overline{f}_2^{\overline{j}} - \tilde{\cal G}_{I \overline{J}} \tilde{f}_1^I \overline{\tilde{f}}_1^{\overline{J}} 
\end{eqnarray}
We have once again omitted several terms that do not depend on any field that we will give a vev in this paper, and thus will not be relevant in what follows.

\vspace{0.2cm}

Given the above structure, One can give a vev to a matter field $C_2$ and satisfy the D-term (\ref{du1}) simply by moving in K\"ahler moduli space to a region where $\mu({\cal F})<0$. Indeed given the lack of $C_1$ fields in this example, all rank preserving deformations of (\ref{mrv3sum}) to an irreducible bundle are of this form \cite{Li:2004hx}. This is, therefore, the field theory description of how the split bundle (\ref{mrv3sum}) is deformed back into (\ref{v3other}) \cite{Anderson:2010ty}.

We can now see that the K\"ahler potential term vanishings that we have shown for $V_3$ can not be due to the effects of this stability wall. It is true that terms of the form $\tilde{f_2} \overline{\tilde{f}}_1$ are forbidden by the Green-Schwarz anomalous $U(1)$ symmetry. However, in returning to the indecomposable bundle, this symmetry has been further broken by the $C_2$ vev that is turned on. In particular, at the stability wall, there is no known symmetry consideration preventing a term of the form $\tilde{f_2} \overline{\tilde{f}}_1 C_2$ appearing in the K\"ahler potential. Once we have a $C_2$ vev turned back on to return to (\ref{v3other}) we should therefore regrow  the $\tilde{f_2} \overline{\tilde{f}}_1 \left<C_2\right>$ term, and its conjugate, if all terms allowed by symmetry are present.

In fact, a moments thought about the statement in the last paragraph reveals that the vanishing of the 12 K\"ahler potential terms that our analysis has revealed for (\ref{v3other}) implies that an {\it infinite} number of terms must have vanished in the matter field K\"ahler potential at the stability wall. For example, terms of the form $\tilde{f_2} \overline{\tilde{f}}_1 C_2 (C_2 \overline{C}_2)^q$ must vanish for all $q$ if $C_2$ is the field to which we give a vev. We will investigate this phenomenon further in our final example in the next section.

\subsection{Vanishing of higher order terms in the matter field K\"ahler potential}

In this section we are going to consider what happens if one takes the example given in Section \ref{eg2} and Higgses by giving an expectation value to a ${\bf 27}-\overline{\bf 27}$ pair. First, let us discuss our expectations from a purely field theory perspective. 

For the matter fields to which we will give a vev we shall denote the decomposition under the breaking of $E_6$ to $SO(10)\times U(1)$ as,
\begin{eqnarray}
\begin{array}{ccccccc}
    {\bf 27}_1 &=& {\bf 1}_{-4,1}&+&{\bf 10}_{2,1}&+&{\bf 16}_{-1,1} \\ 
    f_2 &=& \phi_2 &+& {\chi}_2 &+& \psi_2
    \end{array}\;,
\end{eqnarray}
and
\begin{eqnarray} \label{ftdecomp}
\begin{array}{ccccccc}
    \overline{\bf 27}_2 &=& {\bf 1}_{4,2}&+&{\bf 10}_{-2,2}&+&\overline{\bf 16}_{1,2} \\ 
    \tilde{f}_1 &=& \tilde{\phi}_1 &+& \tilde{\chi}_1 &+& {\tilde{\psi}}_1
    \end{array}\;.
\end{eqnarray}
In these expressions, the first charge is that of the $U(1)$ inside $E_6$ and the second is that of the Green-Schwarz anomalous symmetry. We are going to consider giving a vev to the $SO(10)$ singlet $\phi_2$ in an $f_2$ multiplet to take the unbroken gauge group from $E_6$ to $SO(10)$. To satisfy the D-term (\ref{d2010}) we will also give a vev to the $SO(10)$ singlet $\tilde{\phi}_1$ in an $\tilde{f}_1$ degree of freedom. In performing this Higgsing one $\tilde{f}_1$ and one $f_2$ will get eaten by the massive gauge bosons/fixed by the D-terms. 

Let us denote the decomposition of $\tilde{f}_2$ under the gauge group breaking as follows.
\begin{eqnarray}
\begin{array}{ccccccc}
    \overline{\bf 27}_{-1} &=& {\bf 1}_{4,-1}&+&{\bf 10}_{-2,-1}&+&\overline{\bf 16}_{1,-1} \\ 
    \tilde{f}_2 &=& \tilde{\phi}_2 &+& \tilde{\chi}_2 &+& {\tilde{\psi}}_2
    \end{array}\;.
\end{eqnarray}
From a field theory perspective, we would expect to no longer have vanishing structure in the K\"ahler potential where terms of the form $\tilde{\psi}_2 \overline{\tilde{\psi}}_1$ do not appear. The reason behind this expectation is that in general the K\"ahler potential of the un-Higgsed theory should contain terms such as $\tilde{f}_2 \overline{\tilde{f}}_1 {{f}}_2 {\tilde{f}}_1$ since they are allowed by gauge invariance.
If we put the vevs into the ${f}_2$ and ${\tilde{f}}_1$ factors, we will regrow the type of quadratic K\"ahler potential term that is missing due to gauge symmetry at the stability wall as $\tilde{\psi}_2 \overline{\tilde{\psi}}_1 < {{\phi}}_2 {\tilde{\phi}}_1>$.

With this field theory expectation in mind, let us return to the geometry of the heterotic compactification to ascertain what actually happens. In \cite{Anderson:2010ty} it was shown that the type of Higgsing being considered here corresponds to a rank changing transition which deforms the bundle $V_3$ given in (\ref{v3other}) as follows. Define a bundle ${\cal U}$ by the following short exact extension sequence.
\begin{eqnarray}
\begin{array}{ccccccccc}
    0& \to &{\cal O}& \to& {\cal U} & \to &{\cal K}& \to& 0 \\ 
    h^0 & & 1 & &1&& 0\\
    h^1 && 0 &&0&&0 \\
    h^2 && 0 &&3 &&4 \\
    h^3 && 1&&0&&0\end{array}
\end{eqnarray}
The $SU(4)$ bundle that we deform to under Higgsing is then given by the following extension.
\begin{eqnarray}
\begin{array}{ccccccccc}
    0& \to &{\cal F}& \to& V_4 & \to &{\cal U}& \to& 0 \\ 
    h^0 & & 0 & &0&& 1\\
    h^1 && 5 &&4&&0 \\
    h^2 && 3 &&6 &&3 \\
    h^3 && 0&&0&&0\end{array}
\end{eqnarray}
Here ${\cal F}$ was defined in equation (\ref{fdef}) and we have once again written the cohomology dimensions of the bundles appearing in the defining sequences under the relevant objects.

Looking at the $\overline{\bf 16}$ matter fields, associated to $H^2(V_4)$ we see that there are now $6$ instead of the $7$ $\overline{\bf 27}$ representations that were present in the $E_6$ theory, as a consequence of the usual Higgsing mechanism. We also note that $3$ of these, the $\tilde{\psi}_2$'s, descend from $H^2({\cal U})$ and $3$, the $\tilde{\psi}_1$'s, from $H^2({\cal F})$. Our analysis of Section \ref{vansec} then tells us that the $\tilde{\psi}_2 \overline{\tilde{\psi}}_1$ terms are still missing from the matter field K\"ahler potential, despite the field theory expectations discussed above. This result implies that an infinite number of matter field K\"ahler potential terms that were allowed by the symmetries of the $E_6$ theory, must have in fact been missing. For example all terms of the form $\tilde{f}_2 \overline{\tilde{f}_1} (f_2 \tilde{f}_1)^n$ for any $n$ must vanish in the $E_6$ theory for this result to hold. These terms can not cancel amongst themselves because they all have different dependence on the singlets which gain a vev (and the moduli of the $E_6$ theory in general) and the vanishing result seen in the $SO(10)$ theory holds everywhere in moduli space.

\vspace{0.2cm}

This result is more of a surprise in some ways than the results detailing vanishings of higher order superpotential couplings that were presented in \cite{Gray:2024xun}. We know that we can often build bundles that are poly-stable and holomorphic over a finite moduli space. These therefore correspond to a good perturbative heterotic supersymmetric Minkowski vacuum for finite values of moduli fields. The only way this can be achieved from a four dimensional perspective, given the criteria that $W=0$ and $\frac{\partial W}{\partial \Phi}=0$ for all fields $\Phi$, is if large sets of higher order couplings in $W$ vanish identically to ensure the presence of a flat direction. The situation is different for the K\"ahler potential. $K$ does not appear in the condition for a supersymmetric Minkowski vacuum at all, and it was not obvious that the theory should give rise to such infinite sets of vanishing couplings. Nevertheless, as the above example shows, this structure does seem to be present, at least in the examples that we have been able to study.

\section{Conclusions} \label{sec:conc}

In this paper we have seen that explicit computations can reveal vanishing of terms in the matter field K\"ahler potential of heterotic theories that we would expect to be present purely from symmetry considerations in the four dimensional effective field theory. These vanishings can not in general be achieved by a simple field transformation from a general K\"ahler potential, as such a change of variables would not respect the holomorphic nature of the $N=1$ fields. The vanishing phenomenon arises from bundle substructure in the form of a bundle construction based upon composing smaller objects to make a larger rank whole. This should be compared to the similar situations seen in the vanishing of superpotential couplings \cite{Braun:2006me,Bouchard:2006dn,Anderson:2010tc,Buchbinder:2014sya,Blesneag:2015pvz,Blesneag:2016yag,Gray:2019tzn,Anderson:2021unr,Anderson:2022kgk,Gray:2024xun} which also arise from substructure of the geometric description of the vacuum in various forms. Although we have considered extension bundles here, the techniques used can clearly be extended to other constructions.

In addition to describing in general one way in which the absence of expected terms in the matter field K\"ahler potential can arise, we have provided several explicit examples. These included cases where the results of our analysis can be verified by other means, examples where no other explanation for these vanishings are currently known and examples where infinite sets of contributions to the perturbative K\"ahler potential can be argued to vanish.

One thing that we should emphasize is that a given bundle can be described in a large number of inequivalent manners. It is sufficient for just one of these plethora of descriptions to fit within our framework for the vanishing results we present to hold. Furthermore, if in a given example different descriptions of a bundle lead to different constraints, one could end up with an extremely restricted K\"ahler potential for the resulting four dimensional theory.
       
It would be nice to generalize recent numerical approaches to computing the matter field K\"ahler potential \cite{Constantin:2025vyt,Butbaia:2024xgj,Berglund:2024uqv,Constantin:2024yxh,Butbaia:2024tje} so that examples such as those considered here could be included. In addition to providing cross checks of the two types of work, it is possible that such numerical analyses could reveal further vanishing structures, that could then be studied analytically. 

An important underlying question that was not addressed in this work is whether the structure in the matter field K\"ahler potential discussed here is related to a symmetry in some currently unknown fashion. The answer to this question would clearly be a key underlying ingredient in our understanding of these heterotic vacua, and we plan to address this line of investigation in future work.

\section*{Acknowledgements}

The author is supported in part by NSF grant PHY-2310588. 


\end{document}